\documentclass[aps, prl, superscriptaddress, showpacs, showkeys, reprint]{revtex4-1}

\usepackage{graphicx}
\usepackage{color}
\usepackage{amsmath}
\usepackage{amssymb}
\usepackage{amsfonts}
\usepackage{bm}
\usepackage{natbib}
\usepackage{epstopdf}

\begin{document}

\title{The Impact of Imperfect Information on Network Attack}
\author{Andrew Melchionna}
\affiliation{Institute for Research in Electronics and Applied Physics, University of Maryland, College Park, Maryland 20742 USA}
\affiliation{University of Rochester, Rochester, New York 14627 USA}
\author{Jesus Caloca}
\affiliation{Institute for Research in Electronics and Applied Physics, University of Maryland, College Park, Maryland 20742 USA}
\affiliation{Boise State University, Boise, Idaho 83725 USA}
\author{Shane Squires}
\affiliation{Institute for Research in Electronics and Applied Physics, University of Maryland, College Park, Maryland 20742 USA}
\author{Thomas M. Antonsen}
\affiliation{Institute for Research in Electronics and Applied Physics, University of Maryland, College Park, Maryland 20742 USA}
\author{Edward Ott}
\affiliation{Institute for Research in Electronics and Applied Physics, University of Maryland, College Park, Maryland 20742 USA}
\author{Michelle Girvan}
\affiliation{Institute for Research in Electronics and Applied Physics, University of Maryland, College Park, Maryland 20742 USA}
\date{\today}

\begin{abstract}
This paper explores the effectiveness of network attack when the attacker has imperfect information about the network. For Erd\H{o}s-R\'enyi networks, we observe that dynamical importance and betweenness centrality-based attacks are surprisingly robust to the presence of a moderate amount of imperfect information and are more effective compared with simpler degree-based attacks even at moderate levels of network information error. In contrast, for scale-free networks the effectiveness of attack is much less degraded by a moderate level of information error. Furthermore, in the Erd\H{o}s-R\'enyi case the effectiveness of network attack is much more degraded by missing links as compared with the same number of false links. 
\end{abstract}

\maketitle

Many complex dynamical processes are supported by networks of interconnections between a large number of individual elements (e.g., epidemics \cite*{tuite2010estimated, pastor2001epidemic,  pastor2002immunization, cohen2003efficient}, cancer spread \cite{pomerance2009effect}, electrical power distribution \cite*{dobson2007complex, wang2009cascade}, etc.). Interventions that seek to degrade \cite*{albert2000error,holme2002attack,motter2002cascade, huang2011robustness,callaway2000network,holme2004efficient} or protect \cite*{holme2004efficient,latora2005vulnerability, schneider2011mitigation,chen2008finding} network connectivity are thus of great interest. In particular, strategies for network attack by node or link removal have been intensively studied. Key issues have been the dependence of the attack effectiveness upon network topology and the strategy for selecting nodes or links for removal. We note, however, that, while such previous studies have predominantly presumed the attacker to have perfect knowledge of the network to be attacked, this is very often not the case. Specifically, networks inferred from measurements typically have false links and miss true links. One might suppose that these errors could very much lower the effectiveness of attack strategies. The purpose of this paper is to address this important issue for the case of node removal attacks of undirected networks (directed networks are treated in the online supplementary material \cite{SupMaterial}).

One example of a network attack problem is an attempt to stop the spread of a disease with a limited number of vaccinations: the people who receive the vaccinations are chosen on the basis of their position in the social network \cite*{albert2000error,holme2002attack, motter2002cascade, huang2011robustness,callaway2000network,holme2004efficient}. Another example is that of deriving gene therapies for cancer. Here the goal is to select those genes whose disabling would most inhibit cancer cell survival and proliferation \footnote{A general source on this topic is the journal Cancer Gene Therapy.}\cite*{pomerance2009effect}. Yet another example is the study of the resilience of the Internet to intentional attack \cite*{albert2000error,callaway2000network}.  The typical attack strategy is to calculate some centrality measure of each node, and to then attack (disable, vaccinate, or remove) those nodes with the highest values of this measure. However, an attacker with imperfect network information will determine values of these centrality measures with some error, and using these would be expected to degrade the effectiveness of his attack. Imperfect network information is ubiquitous in applications and can arise in various ways. Examples of link errors can be found in online social networks, where a friendship may be indicated despite the two subjects having never personally met, or inversely, if no online friendship exists between two face-to-face friends. In the previously cited example of cancer gene therapy, genes are selected for disabling based upon an estimated gene interaction network inferred from noisy measurements (e.g., measurements of gene expression \cite*{faith2007large, ku2012interpreting}). Recently, Platig et al.\ studied the effects of link errors on the correlation between network centrality measures inferred from true and erroneous network information \cite{platig2013robustness}. 
\begin{figure*}[t]
\includegraphics[width = 7.0in]{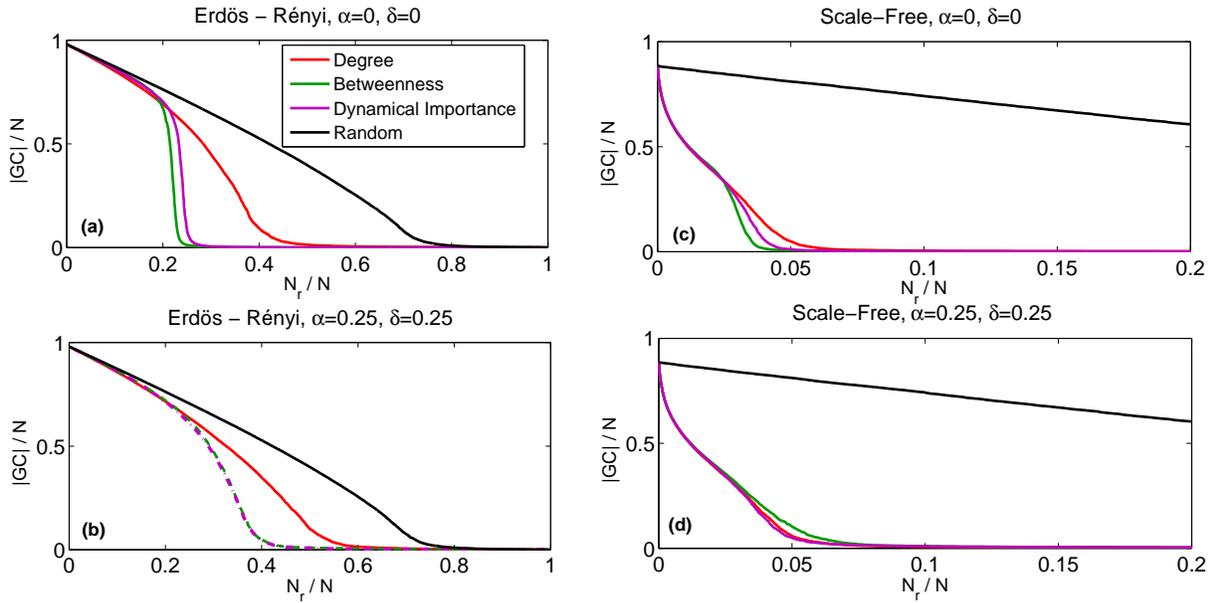}
\caption{The size of the GC normalized by the number of nodes $N$ versus the normalized number of nodes removed $N_\text{r}/N$, for betweenness, dynamical importance, degree, and random strategies for undirected Erd\H{o}s-R\'enyi and scale-free networks, both shown with no error, and with $\alpha=\delta=0.25$}
\end{figure*}

One conclusion of past work for the case where the network is exactly known is that a strategy based on the globally dependent node centrality measure of betweenness (defined subsequently) is particularly effective \cite{holme2002attack}. On the other hand, one might suspect that more effective globally-based strategies are also less robust to error in network knowledge. Our main conclusions are as follows: (i) for Erd\H{o}s-R\'enyi networks, strategies based on global information are surprisingly robust and maintain a clear advantage over the simple node degree-based attack up to moderate amounts of network error; (ii) scale-free networks display much less dependence on the attack strategy (for strategies based on sensibly chosen centrality measures) and much less degradation of attacks by network information error; (iii) for  Erd\H{o}s-R\'enyi networks attack effectiveness is degraded much more by missing links as compared with the same number of false links; (iv) comparing the two global strategies that we test, namely betweenness \cite*{freeman1977set, wasserman1994social} and dynamical importance \cite{restrepo2006characterizing},  betweenness is often slightly more effective at low network error (at the expense of substantially greater computational cost), but the two tend to perform more equally at moderate network error or a relatively small number of attacked nodes; and (v) as shown in the supplemental material, results (i)-(iv) demonstrated in this paper for undirected networks also apply to directed networks. We next describe the numerical experiments that yield  results (i)-(iv). 

\emph{Network Models}. For our ``true'' networks, we consider two types of random networks: Erd\H{o}s-R\'enyi, in which the degree (number of links to a node) has a binomial distribution, and scale-free \cite{barabasi1999emergence}, in which the degree distribution obeys a power law: 
\begin{equation*}
P_k=\dfrac{k^{-\gamma}}{\sum\limits_{i=1}^{k_\text{max}} k_i^{-\gamma}}
\end{equation*}
where $P_k$ is the probability that a randomly chosen node has degree $k$.

\emph{Centrality Measures}. The three node centrality measures upon which we base attack strategies are as follows:
\begin{enumerate}
\item[(i)]The \emph{degree centrality}, which is simply the degree of a node.

\item[(ii)]The \emph{betweenness centrality} of a node is the fraction of shortest paths between all node pairs that pass through that particular node. Let $\sigma(s,t)$ be the number of shortest paths between nodes $s$ and $t$, and $\sigma_i(s,t)$ to be the number of shortest paths between $s$ and $t$ that pass through node $i$. The betweenness of node $i$ is 
\begin{equation*}
b_i = \sum\limits_{s,t,s\neq t\neq i} \dfrac{\sigma_i(s,t)}{\sigma(s,t)}
\end{equation*}
\item[(iii)]The \emph{dynamical importance} of a node is a measure of the change in the largest eigenvalue of the adjacency matrix (which is typically real and positive) upon removal of that node. For an undirected network, elements of the adjacency matrix $A$ are $A_{ij}=A_{ij}=1$ if there is a link between nodes $j$ and $i$, and $A_{ij}=A_{ji}=0$ otherwise. Let $\lambda$ denote the largest eigenvalue of $A$, so that $A\bm{v} = \lambda\bm{v}$ for the corresponding eigenvector $\bm{v}$. Upon removing a node $s$ from the network, and consequently deleting all links attached to it, the matrix $A$ is changed by setting all the matrix elements in row $s$ and column $s$ to zero ($A_{st}=A_{ts}=0$ for all $t$). We use $\Delta\lambda_s$ to denote the resultant change in $\lambda$. The dynamical importance of the node $s$ is defined as 
\begin{equation*}
d_s=-\dfrac{\Delta\lambda_s}{\lambda}
\end{equation*}
with $\lambda$ being the eigenvalue of the matrix before removal of node $s$.
\end{enumerate}
\begin{figure*}[t]
\includegraphics[width = 7in]{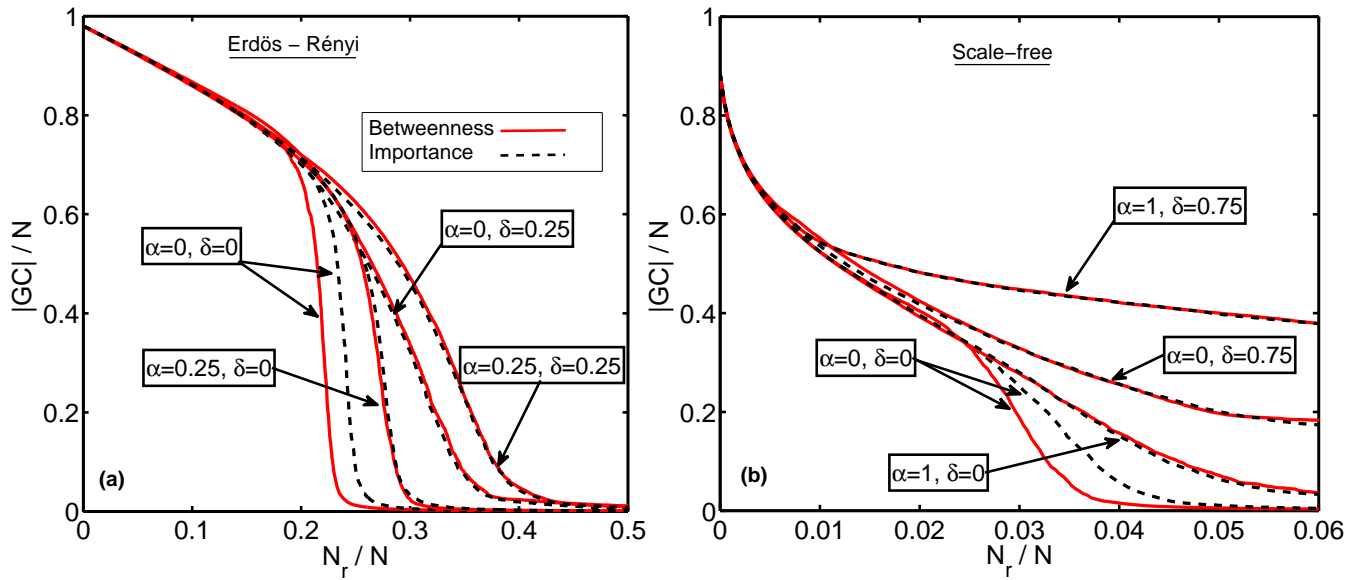}
\caption{The normalized GC size versus the normalized number of removed nodes for betweenness and dynamical importance strategies for undirected (a) Erd\H{o}s-R\'enyi  and (b) scale-free networks.}
\end{figure*}

\emph{``Noisy'' Network Model.} We generate ``noisy'' networks from the true networks by adding false links to the system and removing true links. (We refer to true links which have been removed as ``missing'' links.) Our method for generating noisy networks is as follows \cite{platig2013robustness}: $m_\text{t}\delta$ links are omitted randomly, where $m_\text{t}$ is the number of links in the true network and $0 \leq \delta \leq 1$. Links are only eligible for omission if they are part of the true network; false links added in the adding process will not be omitted. While each link has an equal probability of being deleted, higher-degree nodes have a higher probability of losing a link, as they have more links. In addition, $m_\text{t}\alpha$ false links are added to the network. False links are placed between random pairs of nodes, provided that a link does not already exist between them in the true network. Overall our false network model is characterized by the two parameters $\delta$ and $\alpha$, respectively representing the error levels associated with missing and false links. 

\emph{Description of Numerical Experiments}. The networks are of size $N=2500$, with the maximum possible degree of a node set at $k_\text{max} = N/2 = 1250$. The Erd\H{o}s-R\'enyi networks have an average degree $z_\text{er}=4$, and the scale-free-degree networks have $\gamma \approx 2.06$ and average degree $z_\text{sf}=4$. In the scale-free networks, we require that the degree of each node is at least 1. The networks are constructed according to the configuration model \cite{newman2003structure}. Next, a noisy network is constructed based on the parameters $\alpha$ and $\delta$. The centrality measure is calculated from the existing noisy network, and the highest-centrality node is removed from both the true and noisy networks. If there is more than one node having the same highest value centrality measure, one of those is chosen randomly for removal. Then, we calculate the size of the Giant Component (GC) in the true network (The GC is the largest collection of nodes such that any pair of nodes in the GC is connected by a path along links.). To reiterate, the idea here is that network attacks are based on the information in the noisy network, but the effects of these attacks are actually felt on the true network. After each removal, we recalculate the centrality measure based on the new noisy network (with the previously attacked node deleted), and remove the highest centrality node from both networks again, and recalculate the GC size. This process is continued until all nodes are deleted. 

\emph{Results}. Here, we present the results of numerical simulations exploring the effects of network information errors on attack. Results are averaged over 50 different network realizations. Figure 1 presents the size of the giant connected component of undirected true networks plotted against the number of nodes removed in attack (both normalized by $N$) for Erd\H{o}s-R\'enyi and Scale-Free networks, both for attacks with perfect information (Figs.\ 1(a) and 1(c)) and for attacks with imperfect information ($\alpha=\delta=0.25$) (Figs.~1(b) and 1(d)).  We plot results for attacks based on our three centrality measures (betweenness, dynamical importance, and degree) and, as a baseline, also include results for the case where nodes are successively removed at random. We see that in the case of the Erd\H{o}s-R\'enyi networks (Figs.~1(a) and 1(b)), the betweenness and dynamical importance strategies are significantly better than the degree and random strategies, even with an additional 25\% false links added, and 25\% of true links deleted. 

Furthermore, we see that the betweenness strategy is slightly more efficient than the dynamical importance strategy in the case of no error, and they become approximately equal when error is present. In the case of the undirected scale-free networks (Figs.~1(c) and 1(d)), we find that the degree attack is relatively insensitive to this moderate amount of error. Restricting attention to reductions of the GC to as low as 10\% of its original size, in contrast with the Erd\H{o}s-R\'enyi case, we see that for the scale-free case there is relatively little difference between the different strategies and relatively a much less dramatic effect of moderate network error. 

Figure 2 shows GC attack curves for undirected networks subjected to betweenness and dynamical importance attacks, with different types of error. Figure 2(a) for  Erd\H{o}s-R\'enyi networks shows that at moderate levels of error, $(\alpha,\delta) = (0.25,0), (0,0.25) $ and $(0.25,0.25)$, attacks are more robust to the addition of false links as compared with the omission of the same number of true links. Again, while for  $(\alpha,\delta) = (0,0)$ betweenness based attack is somewhat more effective than dynamical importance based attack, this difference essentially disappears when either of the moderate error types shown are present.

Since Figs.~1(c,d) showed quite weak effects of moderate network error $(\alpha,\delta) = (0.25,0.25)$ for scale-free networks, we are lead to consider substantially higher levels of network error for the scale-free case. Consequently, in Fig.~2(b) we show results for scale-free networks with $(\alpha,\delta) = (0,0),(0,0.75),(1,0),(1,0.75)$ (note that $\alpha = 1$ means that the number of added false links is the same as the number of true links). Even at these high network error values, we find little effect of network error for reductions of the GC size by up to 0.5. For greater reductions of the GC size network error becomes significant, but very great GC reductions are still achieved at relatively small $N_\text{r} / N$ (compared with the Erd\H{o}s-R\'enyi case). The effective absence of network error impact for scale-free networks and $|\text{GC}| / N \gtrsim 0.5 $ can be understood on the basis that reductions of GC size in this range are achieved by removal of a relatively small number of nodes that have extraordinarily high betweenness and importance centrality measures. Random addition of false links, even if it doubles the number of perceived links is unlikely to produce any nodes with centrality measures as high as the true hubs, which will hence still be highly ranked for attack. On the other hand, random deletions with $\delta = 0.75$, on average reduces the degrees of all nodes, roughly proportionally, and assuming connectivity is still maintained between the true hubs, they will still by highly ranked for removal.

In conclusion, we have investigated the impact of imperfect network information on the effectiveness of nodal attack based on different centrality measures (degree, betweenness, and importance). Our results indicate strong dependence on the network degree distribution and on whether the network error is through false links or through missing true links. One implication of the latter finding is that, in the absence of hubs, network inference from noisy data (as in the cancer gene therapy application referred to at the beginning of the paper) should employ a somewhat weaker threshold for link inference (in order to favor inclusion of true links at the possible expense of the addition of false links in the inferred network). There are many possible future extensions of this general line of study, such as investigation of link attacks, the impacts of other network topological characteristics beyond degree distribution (e.g., assortativity by degree \cite{newman2002assortative}, community structure \cite{girvan2002community}, small worldness \cite{watts1998collective}, motifs \cite{milo2002network}, network hierarchical topology \cite{clauset2008hierarchical}, and multilayer structure \cite{de2013mathematical}), considerations of network error in formulating attacks tailored to disruption of specific dynamical processes (e.g., epidemic spread), etc. 

This work was supported by the National Science Foundation under grant PHY-1156454 and by the Army Research Office under grant W911NF-12-1-0101.

\bibliographystyle{apsrev4-1}

\clearpage

\onecolumngrid 

\begin{center}
{\bf \large Supplement for ``The Impact of Imperfect Information on Network Attack''}
\end{center}
\vspace{.2in}

\twocolumngrid

Our paper presented results, discussion and conclusions on the impact of imperfect information on network attack restricted to the case of undirected networks. Here in Figs.~S-1 and S-2 we give results for directed networks. These figures are analogous to Figs.~1 and 2 of our paper that were for undirected networks. To accommodate directedness there are two new aspects of Figs.~S-1 and S-2 as compared with Figs.~1 and 2: (i) the curves for attack based on degree centrality in Figs.~1 and 2 are now each replaced by two curves, one for in-degree-based attack, and one for out-degree-based attack; and (ii) the vertical axes in Figs.~S-1 and S-2 are the normalized size of the Giant Strongly Connected Component (GSCC) rather than the GC of Figs.~1 and 2 (the GSCC is the largest collection of nodes such that for each pair $(i, j)$ of nodes in the GSCC there is a directed path along links both from $i$ to $j$ and from $j$ to $i$). The parameters and degree distributions used for Figs.~S-1 and S-2 are similar to those for Figs.~1 and 2: $N=2500$, $k_\text{max}^\text{in}=k_\text{max}^\text{out}=N/2=1250$ (where $k^\text{in}$ and $k^\text{out}$ denote in-degree and out-degree). The average in-degree and out-degrees for all networks are 4, and, for scale-free networks the in-degree and out-degree distributions are the same with power-law exponent $\gamma\approx2$ for both.

Examination of Figs.~S-1 and S-2 shows that the in-degree and out-degree strategies yield similar results. Furthermore, and most importantly, all of our main general results for undirected networks (points (i)-(iv) at the end of the third paragraph of our paper) are seen to apply to directed networks. 

\begin{figure*}
\setcounter{figure}{0}
\makeatletter 
\renewcommand{\thefigure}{S-\arabic{figure}}
\includegraphics[width = 7in]{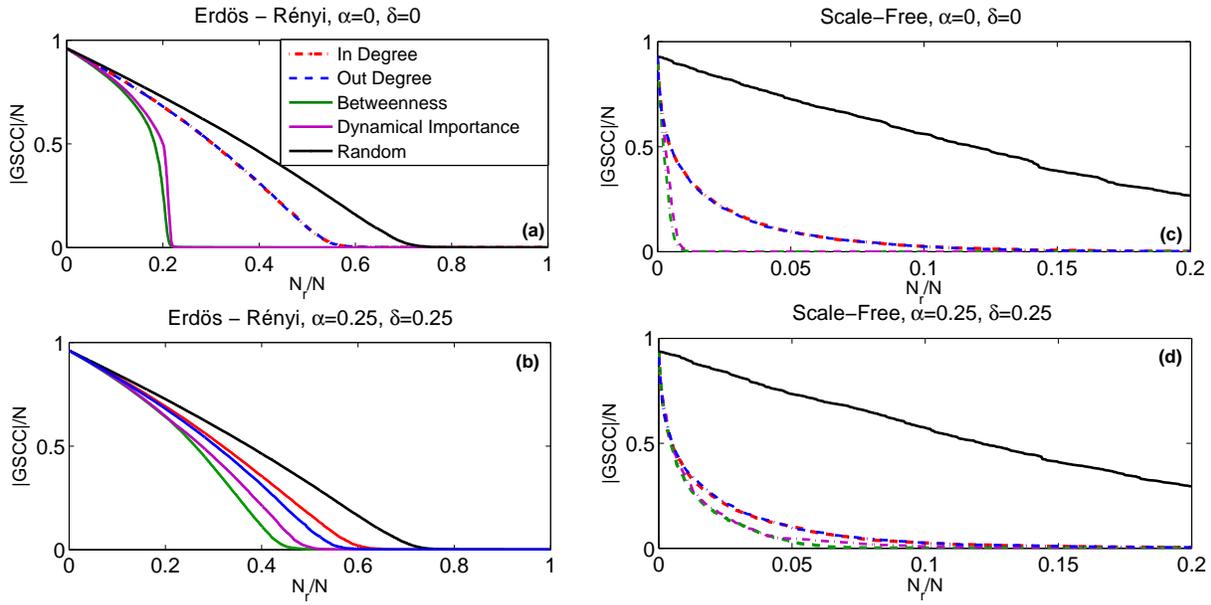}
\caption{\label{FigS1}The size of the GSCC normalized by the number of nodes $N$ versus the normalized number of nodes removed $N_\text{r}/N$, for betweenness, dynamical importance, degree, and random strategies for directed Erd\H{o}s-R\'enyi and scale-free networks, both shown with no error, and with $\alpha=\delta=0.25$.}
\end{figure*}

\begin{figure*}
\makeatletter 
\renewcommand{\thefigure}{S-\arabic{figure}}
\includegraphics[width = 7in]{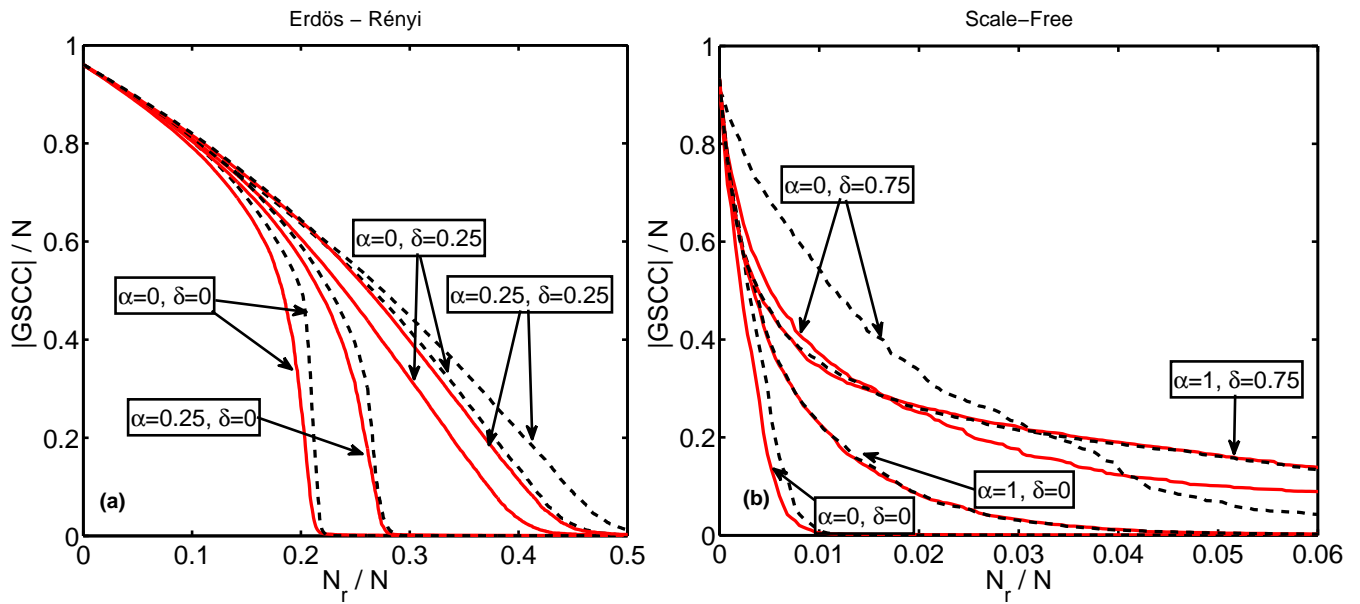}
\caption{\label{FigS2}The normalized GSCC size versus the normalized number of removed nodes for betweenness (solid red line) and dynamical importance (black dashed line) strategies for directed (a) Erd\H{o}s-R\'enyi  and (b) scale-free networks.}
\end{figure*}

\end{document}